\documentclass[structabstract]{aa}  
\usepackage{graphicx}
\usepackage{txfonts}
\usepackage{hyperref}
\hypersetup{
    final=true,
    pageanchor=true,
    colorlinks=true,
    breaklinks=true,
    linkcolor=blue,
    citecolor=blue,
    urlcolor=blue,
    pdfpagemode=UseNone,
    pdftitle={An AR filament studied simultaneously in the chromosphere and photosphere. II. Doppler velocities},
    pdfauthor={C. Kuckein, V. Mart\'\i nez Pillet and R. Centeno},
    pdfsubject={Solar Physics},
    pdfkeywords={Sun: filaments, prominences, 
		Sun: faculae, plages,
		Sun: photosphere,
		Sun: chromosphere,
                Techniques: polarimetric}}

\usepackage{multirow}
\usepackage{natbib}
\bibpunct{(}{)}{;}{a}{}{,} 

\begin{document}

   \title{An active region filament studied simultaneously in the chromosphere and photosphere. 
          II. Doppler velocities}

   \author{C. Kuckein\inst{1,2}
    \and V. Mart\'\i nez Pillet\inst{1}
    \and R. Centeno\inst{3}}   
  
   \institute{Instituto de Astrof\'\i sica de Canarias, V\'\i a 
             L\'{a}ctea s/n, E-38205 La Laguna, Tenerife, Spain\\
             \email{ckuckein@iac.es}
   \and Departamento de Astrof\'\i sica, Universidad de La Laguna, E-38206 La Laguna, Tenerife, Spain
   \and High Altitude Observatory (NCAR), Boulder, CO 80301, USA}

  \date{Received 25 January 2012 / Accepted}

 
  \abstract
   {Paper I presents the magnetic structure, inferred for the
     photosphere and the chromosphere, 
   of a filament that developed in active region (AR) NOAA 10781,
   observed on 2005 July 3 and July 5.}
   {In this paper we complement those results with the 
   velocities retrieved from Doppler shifts measured at the chromosphere and the photosphere in the AR 
   filament area.}
   {The velocities and magnetic field parameters were inferred from full Stokes inversions 
    of the photospheric \ion{Si}{i} 10827\,\AA\ line and the chromospheric \ion{He}{i} 10830\,\AA\ triplet.
    Various inversion methods with different numbers of
      atmospheric components and different weighting schemes of the Stokes 
    profiles were used. The velocities were calibrated on an absolute scale.}
   {A ubiquitous chromospheric downflow is found in the faculae surrounding the filament, with an average
   velocity of $1.6$\,km\,s$^{-1}$. The filament region, however,
   displays upflows in the photosphere on both days,
   when the linear polarization (which samples the transverse component
   of the fields) is given more weight in the
   inversions. The upflow speeds of the transverse fields in the
 filament region average $-0.15$\,km\,s$^{-1}$.
   In the chromosphere, the situation is different for the two days of observation.
   On July 3, the chromospheric portion of the filament is moving upwards as a whole with a mean 
   speed of $-0.24$\,km\,s$^{-1}$. However, on July 5
   only the section above
   an orphan penumbra shows localized upflow patches, while the rest of the filament is dominated by the same 
   downflows observed elsewhere in the facular region.
   Photospheric supersonic downflows that last for tens of
     minutes are detected below the filament, close to the PIL.}
   {The observed velocity pattern in this AR filament strongly
     suggests a scenario where the transverse fields 
   are mostly dominated by upflows. The filament flux rope is seen to be emerging
  at all places and both heights, with a few exceptions in the
  chromosphere. This happens within a surrounding facular
   region that displays a generalized downflow in the chromosphere
   and localized 
   downflows of supersonic character at the photosphere. 
   No large scale downflow of transverse field lines is observed
   at the photosphere.
     }

   \authorrunning{Kuckein et al.}
   \titlerunning{Simultaneous study of the Doppler velocities in an AR filament}
   \maketitle

\section{Introduction}
Velocity measurements in active region (AR) filaments are extremely scarce in
the literature. Most studies have  been carried out only in quiescent (QS)
filaments (outside ARs). However, quantifying velocities and plasma flows
in filaments is a
fundamental step towards understanding their formation, evolution and disappearance.
Filaments are structures formed from cool dense plasma that typically lies
between opposite polarities,
i.e., at the polarity inversion line \citep[PIL;][]{babcock55}, in the
chromosphere or corona. Due to the temperature contrast with their surroundings, they
appear as dark structures when observed on the solar disk.  
Filaments are called prominences if observed in
emission above the limb, but  both terms are often used interchangeably
in the literature. 

To date, there are two proposed scenarios for the formation process of
filaments: the sheared arcade (SA) model and the flux
rope emergence (FRE) model. Both result in very similar global structures,
i.e., a flux rope that hangs
above the photosphere (although the SA model combines dipped field lines
and flux ropes). However, it is in the formation process of the
flux ropes where the two models differ. 
The SA model requires shearing motions of the footpoints to form the flux rope in the corona 
with the aid of reconnection processes identified in the photosphere as cancellation events.
On the other hand, in the FRE model the flux rope is
formed deep in the convection zone (CZ) before it emerges
through the surface and makes its way up into the corona.
However, simulations based on buoyancy instabilities encounter difficulties
when lifting up a mass-loaded flux rope from the photosphere into the corona
\citep[e.g.,][]{murray06,mactaggart10}.  \citet{PaperI} (hereafter Paper I)
provide a more thorough discussion and additional references to these two models
\citep[see also the extensive review by][]{mackay10}.  

Nowadays, high-resolution imaging has shown that quiescent
filaments are made of smaller-scale structures
\citep[e.g.,][and references
therein]{demoulin87,lin05}.  Doppler measurements in these quiescent
filaments \citep[e.g.,][]{marters81,
schmieder10} have revealed that a wide range of velocities coexist in the same
structure. For example, the recent study by \citet{chae07} reported
line-of-sight (LOS) velocities between $\pm 5$\,km\,s$^{-1}$ and $\pm
15$\,km\,s$^{-1}$ in the spine and in threads (respectively) of a QS filament
(at coordinates N15-E26). \citet{mein77} found that the axis of an AR filament
(at N7-W30) was at rest, while downflows on one side and upflows on the other
side were detected. 
Upward motions have also been detected below QS and
AR filaments by \citet{ioshpa99}.
However, the main limitation of Doppler line shifts, and
hence of the retrieved LOS velocities, is that the
inferred values are position-dependent
all over the solar disk. Therefore, it is hard to compare Doppler velocities
measured in different filaments. In addition, a thorough calibration, i.e., 
setting a correct zero value for the rest wavelength, is necessary, especially
when calculating photospheric velocities (below the filament) which are much
smaller \citep[$\leq 0.3$\,km\,s$^{-1}$;][]{martres76} than those
found in the chromosphere and corona \citep[see][]{welsch12}. 

It is presently unclear whether there is an unequivocal way of distinguishing
which of the different filament formation scenarios applies to the
observed AR filament. In this paper we propose to
shed some light on this topic by studying the observed Doppler velocities
inside the AR filament that was analyzed in Paper I.  
In particular, we can ask what the
expected velocity patterns (in and below the filaments) associated with these
two models are.  The flux rope models proposed by
\citet{vanballe89} and \citet{vanballe07} are built through successive
photospheric flux cancellations driven by motions that force the
field lines to converge towards the PIL. After the oppositely directed
field lines are brought together, reconnection processes above the photosphere
create a flux rope in the corona.  The authors
suggest that transverse magnetic field lines (loops which are almost
perpendicular to the PIL) submerge at the PIL, and are transported
downwards into the CZ as a result of the reconnection processes
mentioned before \citep[see panels d and e in Fig.\ 1 of][]{vanballe89}. The
typical size of these submerging loops is small ($\sim$\,900 km),
which is large enough to be detected with currently available
instruments.  

The prominence model presented in the work of \citet{DeVore00}
does not show convergence of flux at the PIL. The 
helicity in the field lines is
built up in the corona as a consequence of reconnection of the highly sheared field
lines (twisted and stretched by the sheared displacements of the footpoints) with the overlying
coronal arcade. The authors point out that the resulting magnetic structure is
stable; hence, no upflows or downflows are expected in this model. However, the observed
evolution of filaments suggests that they are of a non-steady
nature. This calls for models with built-in evolutionary processes
that allow the filament to disappear from the solar surface,
transporting magnetic fields upwards or downwards through the
photospheric boundary.

The first complete description of the emergence of a flux rope AR filament was presented 
by \citet{okamoto08,okamoto09}.  The authors show observational evidence of an
emerging helical flux rope that apparently reconnects with the remnant
magnetic field of a pre-existing filament. 
A Milne--Eddington (ME) inversion of the Stokes profiles observed with
Hinode/SP over the region where the transverse fields of the emerging flux rope were 
observed yielded a mean upflow of $-0.3 \pm 0.2$\,km\,s$^{-1}$. The ME inversion 
applied by these authors was able to separate the velocity of the
magnetic and the non-magnetic components (see their Fig.\ 2).
Interestingly, while the non-magnetic atmosphere showed a typical
granulation pattern, upflows
were systematically inferred for the {\em magnetic}
component. Thus, in that work, the picture of the filament formation is compatible
with the existence of transverse fields that are moving upwards into the chromosphere
and corona.
\citet{mactaggart10} simulated the rise of a weakly twisted flux tube from
the solar interior up into an overlying arcade in the corona. The axis of their
flux rope became trapped slightly above the photosphere, while the field lines
of the top of the flux rope were still able to reach coronal heights. In the
same way, the 3D MHD simulations of an emerging flux rope presented by
\citet{yelles09} showed horizontal field lines that moved upwards with
velocities of $\sim - 0.67$\,km\,s$^{-1}$. However, the authors pointed out
that the emergent flux studied by \citet{okamoto09} was 
probably less buoyant than the case simulated in their work, since the observed rising
speed was smaller and the modification of the granular pattern less severe
than in the simulations. Interestingly, \citet{lites10} also found a
similar pattern to that presented by \citet{okamoto09}
in their study of the evolution of a filament channel inside an active
region. The inferred magnetic structure in the
filament channel corresponded to horizontal field lines elongated along the PIL 
(with equipartition strengths) that often (but not always) displayed
upflows \citep[see the $v_{mag}$ panel in Fig.\ 5 of][]{lites10}. 
A slow continuous upward motion in active region filaments had already been
observed almost three decades ago \citep{malherbe83,schmieder85}, albeit
in chromospheric layers. It is thus clear that studying the behavior 
of transverse magnetic fields at the location of filaments (i.e., at AR PILs),
is crucial to understanding their evolution \citep[see also the discussion
in][]{welsch12}.
In particular, it will help to shed some light on how the magnetic structures that 
sustain the filaments get filled with plasma and how they eventually get rid of
it, unloading most of their mass. To this end, it is crucial to obtain more data
of the flows associated with filament formation and evolution. 

In Paper I, we presented the magnetic structure of an AR filament, inferred from simultaneous
spectropolarimetric observations in the photosphere and
chromosphere. The filament seemed to be divided into
two parts: (1) one section that showed a flux rope configuration which had
its main axis lying in the chromosphere; and (2)  another section,
which was likely
to have a similar magnetic structure, but whose main axis was detected
in the photosphere instead. This extremely low-lying part of the filament 
left an imprint on the photosphere in the shape of an 
orphan penumbral system in the proximity of small pore
associations. Paper I suggested  that these features were the photospheric counterpart of the filament.
However, in order to understand the formation mechanism of
this AR filament, a detailed study of the Doppler velocities, in a global
scenario, needs to be carried out. The present paper addresses
this issue by focusing in particular
on the photospheric velocities inferred from the Doppler shifts of the \ion{Si}{i}
line in order to search for systematic upflows and downflows. We also analyze the chromospheric 
Doppler shifts provided by the \ion{He}{i} triplet.

\section{Observations}

The active region under study, NOAA AR 10781, was in its slow decay phase 
during our observing run in July 2005. The active region had a round leader sunspot of positive 
polarity and was followed by an extensive plage region that stretched towards higher latitudes. 
On top of the polarity inversion
line (PIL), a filament with an inverse S-shape was unambiguously
identified. Two spectropolarimetric maps
were taken on July 3 (at latitude and longitude N16-E8, $\mu \sim
0.95$), and seven on July 5 (at N16-W18, $\mu \sim 0.91$), using the
Tenerife Infrared Polarimeter \citep[TIP-II;][]{tip2} at the German Vacuum
Tower Telescope (VTT) on Tenerife.  In addition, a time series of $\sim 19$
minutes, with the slit fixed at the PIL, was acquired on July 5.

  \begin{figure}[!t]
   \resizebox{\hsize}{!}{\includegraphics{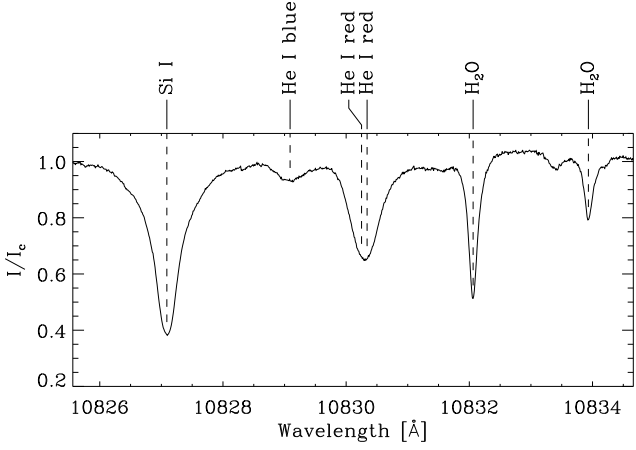}}
    \caption{Stokes $I$ profile, normalized to the continuum intensity, observed by TIP-II 
     near the polarity inversion line. The dashed vertical lines indicate the positions of the rest wavelengths 
    for the spectral lines that are used in this work.}
    \label{Fig:tipspectra}
    \end{figure}

The slit (0\farcs5 wide and 35\arcsec\ long, with a pixel size of 0\farcs17
along the slit) was aligned with the filament and a series of spectropolarimetric maps, centered on the filament, were acquired
using a scanning step size of  0\farcs4  or 0\farcs3, depending on the seeing conditions.
Figure \ref{Fig:tipspectra} shows the spectral range of the spectropolarimeter, which
comprises the photospheric \ion{Si}{i} 10827\,\AA\ line, the chromospheric
\ion{He}{i} 10830\,\AA\ triplet and two telluric H$_2$O lines. The helium
triplet comprises a ``blue'' component and two blended ``red''
components. Table \ref{table:wavelengths} provides the wavelength
and the logarithm of the oscillator strength times the multiplicity of the lower level,
$\log gf$, for each line. This unique spectral window allows us to study the
photosphere and the chromosphere simultaneously. The seeing conditions 
during the observations were highly variable and the adaptive optics system of the VTT \citep[KAOS;][]{kaos}
helped to substantially improve the image quality. The data sets were corrected
for flat field, dark current and polarimetric calibration using
the standard procedures for TIP \citep{collados99,collados03}.  

  \begin{figure}[!t]
   \resizebox{\hsize}{!}{\includegraphics{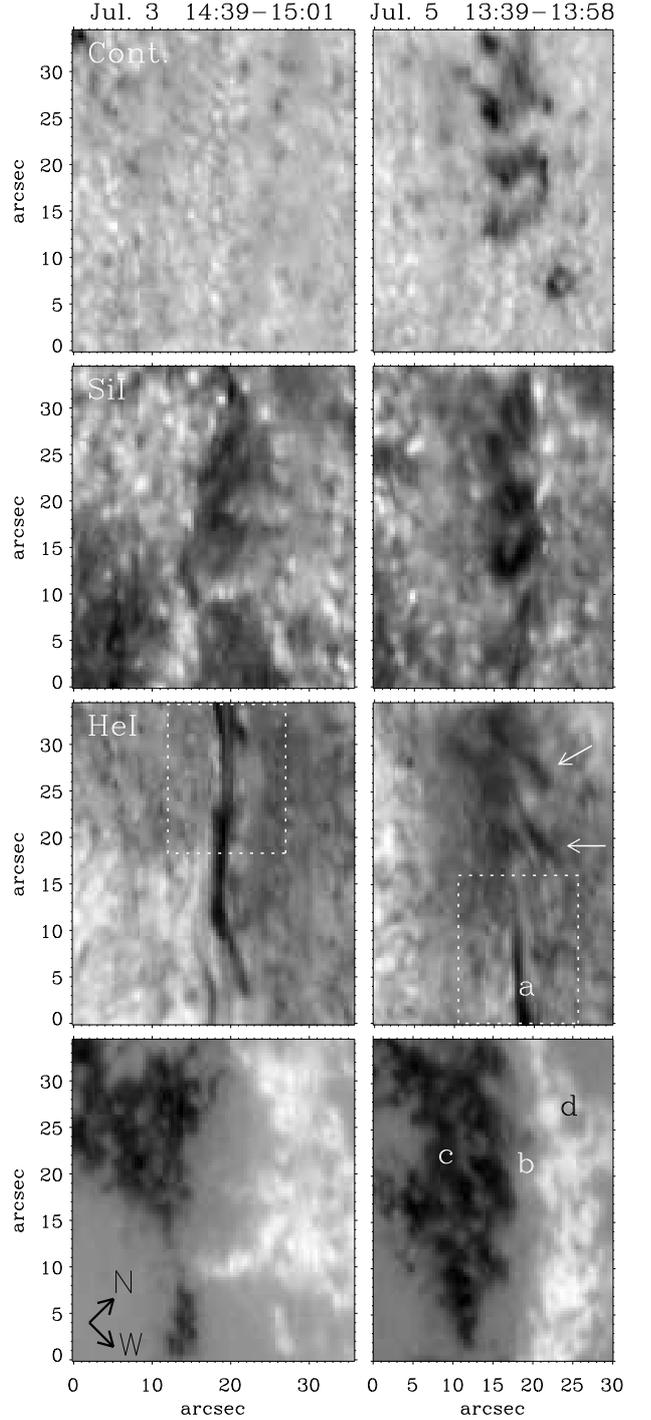}}
    \caption{TIP-II slit-reconstructed images for both days of observation. 
  From \textit{top} to \textit{bottom}: continuum, \ion{Si}{i} line core
  intensity, \ion{He}{i} line core intensity (red component) and \ion{Si}{i} Stokes $V$. The FOV was different on 
  July 3 and July 5. We provide the dashed rectangles to show the approximate common area of the 
filament on both days. Characters a--d indicate the position of the Stokes profiles presented in Fig.\ 
\ref{Fig:coronalrain}. The white arrows point to high concentrations of helium with elongated shapes 
(thread-like structures).}
    \label{Fig:TIP}
    \end{figure}

Figure \ref{Fig:TIP} shows an example, for both days, of the scanned area of the
filament. From \textit{top} to \textit{bottom}, the slit-reconstructed images presented correspond to different
wavelength bands: continuum intensity,
\ion{Si}{i} line core intensity, \ion{He}{i} line core intensity centered at 
the red component and \ion{Si}{i} line-of-sight magnetogram.  On July 3, the 
filament had a thin elongated  shape in the chromosphere (see \ion{He}{i} panel of Fig.\ 
\ref{Fig:TIP}), which we interpret to be the main axis or spine of the filament.
Underneath, in the photosphere, 
the continuum image shows no outstanding features. 
The FOV of July 5 was slightly different and included newly appearing pores and 
orphan penumbral structures in the photosphere, mainly located
along the PIL and below the filament (see continuum image in the
\textit{top righthand} panel). The white dashed
rectangles in Fig.\ \ref{Fig:TIP} show the approximate portion of the filament
 common to both days of observation. The \ion{He}{i} intensity
image of July 5 also
shows the spine of the filament in the lower (common) part of the FOV. This
feature is traced by an elongated shape in the \ion{Si}{i} line core image,  
indicating that the structure lies at a lower mean height than on July 3.
In the upper part of the \ion{He}{i} image, on the other hand, the filament is more diffuse and less
compact, and shows a series of dark threads that are highlighted by the white arrows 
(see also Fig.\ 5 of Paper I for more TIP-II images from July 5). Since the structure of the filament
is identifiable in the silicon images, Paper I concluded that the filament was
extremely low-lying and was directly responsible for the orphan
penumbral aggregations and pores. 

The main purpose of this paper is to identify what the plasma motions
associated with the filament are, both in the chromosphere and the photosphere. In particular, we
concentrate on whether the field lines are moving upwards, which would
indicate an emerging process, or downwards, which would reveal a submergence phenomenon. The
position on the solar disk of this active region is quite close
to disk center on both days. This presents an advantage for the study of line-of-sight
(LOS) velocities, which should not differ too much from the vertical velocities
(measured in a local solar reference frame). For a better understanding of
the inferred velocities we will distinguish between two regions in the filament: the
spine area and the diffused filament area (above the orphan penumbrae),
the latter being observed
only on July 5. As stated in Paper I, these two areas
correspond to the same filament, but observed at different heights. 
It is therefore helpful to study them separately.  

A detailed analysis of the evolution of this AR was presented in Paper I. 

\begin{table}
\caption{Atomic data used in this work.}              
\label{table:wavelengths}      
\centering                                      
\begin{tabular}{ccc}          
\hline\hline                        
 Line & Wavelength (\AA) & $\log gf$ \\    
\hline                                   
    \ion{Si}{i}      & 10827.089\tablefootmark{a} & 0.363$^\star$ \\      
    \ion{He}{i} blue & 10829.091\tablefootmark{a} & $-$0.745$^\dagger$ \\
    \ion{He}{i} red  & 10830.250\tablefootmark{a} & $-$0.268$^\dagger$ \\
                     & 10830.340\tablefootmark{a} & $-$0.047$^\dagger$ \\
    H$_2$O (telluric)& 10832.108\tablefootmark{b} & -- \\
\hline                                             
\end{tabular}

\tablefoot{Wavelengths \tablefoottext{a}{from the NIST database
}
\tablefoottext{b}{inferred in this work; see Appendix \ref{app:telluric}.}
}

\tablebib{
($\star$)~\citet{borrero03}; ($\dagger$) VALD; \citet{VALD99}
}
\end{table}

\section{Data analysis}
To infer the physical parameters from the available sets of spectropolarimetric 
data we used two different inversion codes. It is important to recall  
that an inversion code provides full vector magnetic fields, 
but only the line-of-sight component of the velocity,
in the observer's reference frame. Since one infers the three
components of the vector magnetic field, the resulting inclinations and
azimuths can be projected onto the local solar reference frame
(i.e., with respect to the solar vertical, $z$, and
along the solar latitude and longitude planes). 
However, difficulties arise when this
transformation is carried out. The well-known 180$^\circ$ ambiguity
(due to the angular dependence of Stokes $Q$ and $U$ with the azimuth)
provides two solutions for the magnetic field, with different
azimuths and inclinations when projected onto the local solar
reference frame. To find the correct solution, we used the AZAM
code for the disambiguation of the magnetic field vector \citep[][]{lites95}
(see also Sect.\ 3.2 of Paper I for a more detailed explanation). 
Doppler shifts  provide only one component of the velocity
vector, the one projected along the LOS, rendering the projection onto the local solar frame impossible.
Since the aim of this work is to compare
the LOS velocities and their associated vector magnetic fields, for
the remainder of this paper we
will leave the magnetic field inclinations and azimuths in the 
observer's frame. 
As our observations were done relatively close to disk center, the
distinction between the two frames is more quantitative than qualitative.
In order to show the impact of using the vector magnetic field in the
observer's frame instead of in the local solar frame,
Table \ref{table:LOSAZAM} provides two examples of the inclinations ($\gamma
\rightarrow \gamma_\odot$) and the azimuths ($\phi \rightarrow
\phi_\odot$), averaged over several pixels,
in both reference frames. An inclination of $\gamma = 90^\circ$ in the LOS frame means that the
field lines are perpendicular to the LOS, while $\gamma_\odot = 90^\circ$
means that they are parallel to the solar surface. 
As can be deduced from Table \ref{table:LOSAZAM}, while the exact values
differ between the two frames, the horizontal or vertical nature of the
field vectors is mostly preserved.

\begin{table}
\caption{Examples of the line-of-sight to local solar reference frame transformation of the magnetic 
field inclination and azimuth. The examples were averaged over 11 (PIL) and
8 (spine) pixels, in addition to the original binning of the data (see Sect. \ref{sect:dataanalysis}).}
\label{table:LOSAZAM}      
\centering                                      
\begin{tabular}{cccccccc}          
\hline\hline                        
      &       &      & \multicolumn{2}{c}{LOS frame} &  & \multicolumn{2}{c}{Solar frame} \\ 
  Day   & Location         & Ion    & $\gamma$ ($^\circ$) & $\phi$ ($^\circ$) &  & $\gamma_\odot$ ($^\circ$) & $\phi_\odot$ ($^\circ$) \\ 
\hline                                   
 \multirow{2}{*}{July 3rd} & \multirow{2}{*}{Spine} &   \ion{He}{i}  & 101.5  & 93.0 & $\rightarrow$ & 116.1 & 85.3  \\ 
                                           &        &   \ion{Si}{i}   & 84.2   & 63.6 & $\rightarrow$ & 98.1  & 54.6 \\
\hline                                   
 \multirow{2}{*}{July 5th} & \multirow{2}{*}{PIL}   &   \ion{He}{i}  & 91.6  & 140.4 & $\rightarrow$ & 74.0 & 125.7  \\     
                                           &        &   \ion{Si}{i}  & 88.9  & 114.6 & $\rightarrow$ & 81.2 & 100.0 \\

\hline                                             
\end{tabular}
\end{table}

\subsection{Helium 10830\,\AA\ and silicon 10827\,\AA\ inversions} \label{sect:dataanalysis}
Before inverting the helium spectra we carried out a $3 \times 6$
binning of the data along the scanning and slit directions, respectively,
and a binning of $3$ pixels in the spectral domain to increase the
signal-to-noise ratio (S/N), rendering a value of $\sim 2000$
for all maps. The resulting spectral sampling is $\sim 33.1$\,m\AA\,px$^{-1}$ and
the pixel size $1.2 \times 1.0$ arcsec$^2$ and $1 \times 1$
arcsec$^2$, for July 3 and 5, respectively. However, since the \ion{Si}{i}
line is much stronger than the  \ion{He}{i} triplet, the inversions of
the former were carried out without any binning, the pixel size being $0.40 \times
0.17$ arcsec$^2$ and $0.30 \times 0.17$ arcsec$^2$, on July 3 and 5
respectively. This preserved the original spatial
resolution at an expense of a lower S/N of $\sim 500$. The binned silicon data and
inversions were  used solely for the purpose of Table \ref{table:LOSAZAM}.

The chromospheric \ion{He}{i} 10830\,\AA\ triplet was inverted using a
Milne--Eddington inversion code \citep[MELANIE;][]{melanie}.
This code assumes that the source function varies linearly with optical depth while
the remaining semi-empirical parameters stay constant with height.
MELANIE computes the Zeeman-induced Stokes profiles in the incomplete Paschen--Back (IPB) effect
regime \citep[see][for a better understanding of the IPB effect on the Stokes profiles in the 
\ion{He}{i} 10830\,\AA\ region]{socas04, sasso06}
and has a set of nine free parameters, which
are iteratively modified by the code in order to obtain the best match
between the synthetic and the observed
Stokes data. The macroturbulence and the filling factor,
or fraction of the magnetic component occupied inside each pixel,
were fixed at 1.2\,km\,s$^{-1}$ and $f = 1$ in the inversion, so no
stray-light was used. 
Atomic-level polarization is not taken into account in MELANIE. However,
this is not necessary for this AR filament, since \citet{kuckein09}
proved that these \ion{He}{i} Stokes profiles are dominated by the 
Zeeman effect.

The photospheric \ion{Si}{i} 10827\,\AA\ line was inverted with the SIR code
\citep[\textit{Stokes Inversions based on Response functions};][]{sir} which
solves the radiative transfer equation assuming local thermodynamic
equilibrium (LTE). 
SIR gives a depth-dependent stratification of the inferred
physical parameter as a function of the
logarithm of the LOS continuum optical depth at 5000\,\AA. 

We carried out two sets of inversions for the \ion{Si}{i}
line. The first set, referred to hereafter as ``standard'' inversions, 
used a uniform weighting scheme for the four Stokes parameters:
$w_{I,Q,U,V} = 1$. An average
intensity profile of the non-magnetic areas of each map (regions
where Stokes $Q$, $U$, and $V$ had negligible values), was used as
representative of the stray-light.
The stray-light fraction, $\alpha$, used here as a proxy for the
filling factor, was left as a free parameter in
the inversion. Note that in the case
of helium, the line is weak in quiet regions, so the use of a magnetic
filling factor does not provide any benefit.
We studied the response function (RF) to velocity perturbations in several
model atmospheres obtained from the inversion of the filament
data, finding a consistent value of $\log \tau = -2$ for the 
height at which the sensitivity was the highest. 
The height stratification derived by the inversion code was not used
since the velocities at other optical depths had larger error bars.

The Doppler shift of the stray-light profile cannot be set as a
free parameter in the SIR inversion code.
This extra freedom was used in the studies by \citet{okamoto09} and
\citet{lites10} to separate the magnetic and non-magnetic Doppler shifts. While
the latter were dominated by granulation patterns, the former provided clear
indications of upflows -- a crucial result that we want to investigate 
further in this paper. To this end, we produced a second set of SIR inversions that 
were able to retrieve the Doppler shifts of the magnetic component without
interference from a possible shift of the stray-light profile. These
inversions, referred to heretofore as the ``magnetic'' inversions, forced
a zero-weighting ($w_I = 0$) for Stokes $I$, basically
eliminating any sensitivity to the intensity profile, and hence to the 
stray-light, in the inversion procedure. 
Furthermore, since we were interested in the Doppler shifts of the
transverse fields near the PIL, we weighted Stokes $Q$ and $U$ equally
with $w_{Q,U} = 1$ and set the weight of Stokes $V$ to $w_V = 0.1$.
The filling factor was fixed to $f = 1$. The remaining parameters were
initialized in the same way as in the ``standard'' silicon inversions.  

\citet{bard08} reported in their work that the \ion{Si}{i}
10827\,\AA\ line is formed in non local thermodynamic equilibrium (NLTE)
conditions. In Paper I we tested the NLTE effects on the inversions of the
\ion{Si}{i} line in the filament, using
the departure coefficients, $\beta$ (defined as the ratio between the
population densities calculated in NLTE and LTE), reported in \citet{bard08}.
We found almost negligible changes in the retrieved LOS
velocities,   temperature being the only parameter
that was substantially affected in the upper layers. Therefore, we decided
not to consider the departure
coefficients for the purpose of the present work.

The reader is referred to Paper I for a more thorough description of the
inversions and NLTE effects. 

\begin{figure*}[!t]
\centering
\includegraphics[width=0.99\textwidth]{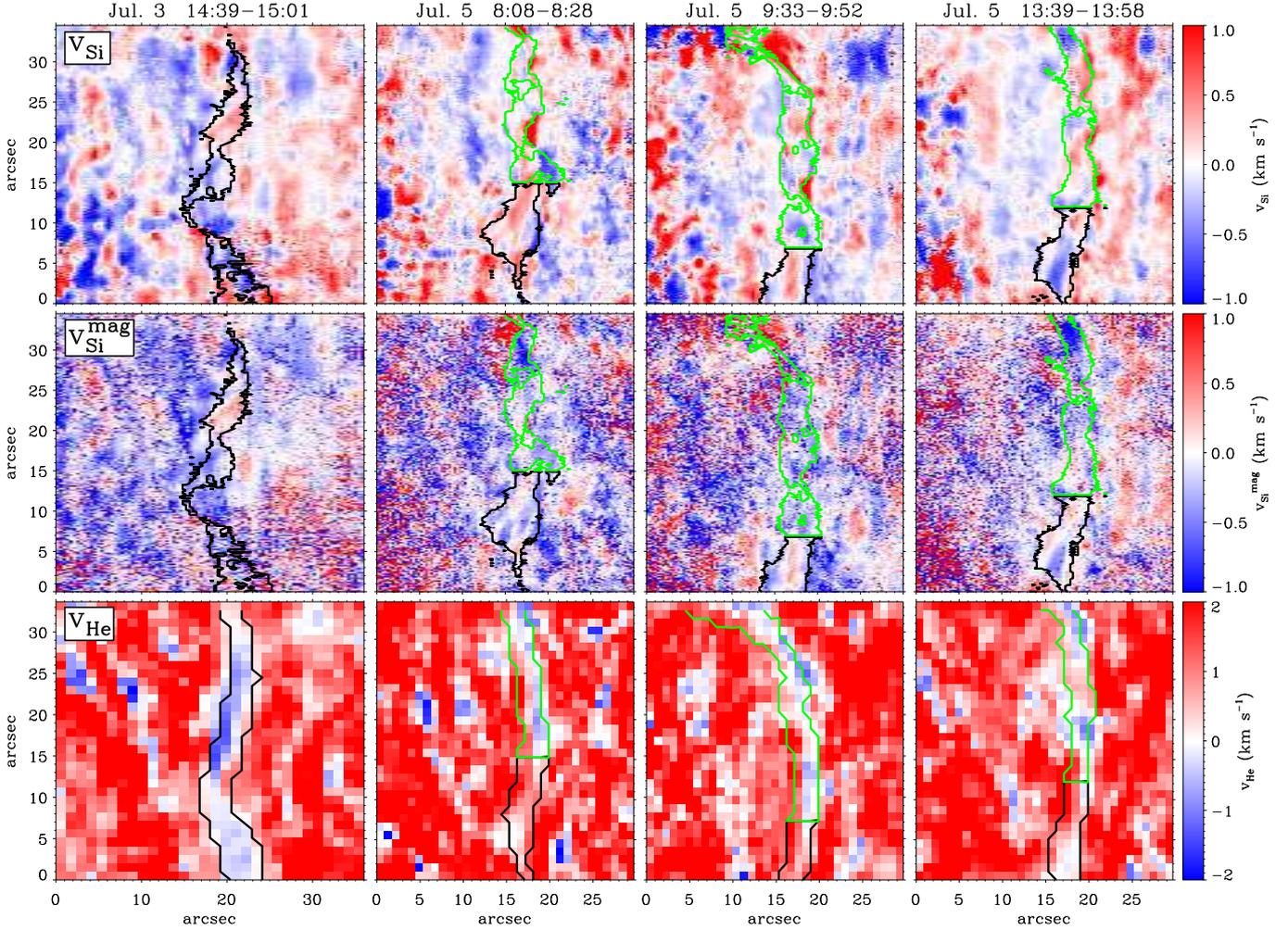}
\caption{Line-of-sight velocities inferred from: the \ion{Si}{i} 10827\AA\ full Stokes ``standard'' 
inversions (\textit{top} row), the
\ion{Si}{i} 10827\AA\ ``magnetic'' inversions (with weights: $w_I= 0$, $w_{Q,U} = 1.0$ and $w_V = 0.1$) 
(\textit{middle} row) and
\ion{He}{i} 10830\AA\ inversions (\textit{bottom} row) at different times. Note that the color scale
is saturated at [$-1$,$1$] and [$-2$,$2$] km\,s$^{-1}$ for silicon and helium respectively. Pixels 
inside the contours have magnetic field inclinations, with respect to
the LOS, between 
$75^\circ \leq \gamma \leq 105^\circ$. In the July 5 maps, the
colored contours distinguish between
the orphan penumbrae area (green) and the spine region (black). On July 3, only the spine of the filament was
seen. Negative velocities (blue) indicate upflows while positive velocities (red) indicate downflows.}
\label{Fig:vmaps}%
\end{figure*}

\subsubsection{Two-component inversions}
The presence of several atypical Stokes $V$ profiles (i.e., with
several highly redshifted lobes) in the \ion{Si}{i} 10827\,\AA\ line led us to use a
two-magnetic component SIR inversion for these cases (see Sect.\ \ref{Sect:supersonicdown}).
Therefore, we had two initial guess atmospheres. 
For the first one, we used the same atmospheric model as that
from the previous single-component inversion, but for the second,
we increased the magnetic field strength from 500
to 1500\,Gauss and the velocity from $0.5$ to 10\,km\,s$^{-1}$. These initial
conditions resulted in a good performance of the SIR code.

\subsection{Wavelength calibration}
The main aim of this work is to provide reliable Doppler velocities measured
at the filament at both heights. In order to calibrate these velocities on
an absolute scale, we had to correct them for several effects such as Earth's
orbital motions and the solar gravitational redshift (see Appendix \ref{app:a} for a
complete description of the calibration). The wavelengths presented in Table
\ref{table:wavelengths} were all obtained from the \textit{National Institute
of Standards and Technology} (NIST) except for the one corresponding to the
telluric line. The literature reports several values,
with differences of the order of tens of m\AA, which are large
enough to cause systematic errors in our velocity calibration.
We propose a new value for the rest wavelength of this telluric
line, which we inferred by calibrating flat field TIP-II data 
taken at disk center (see Appendix \ref{app:telluric}).

\section{Results}

\subsection{Doppler velocities}

 \begin{table*}[!ht]
 \caption{Mean Doppler velocities ($<v>$), of all available maps for both days, calculated inside the contours, 
where the inferred LOS inclinations are between 75$^\circ < \gamma < 105^\circ$, i.e., along the PIL.
We distinguish between two different areas: the filament axis or 
spine (black contours in Fig.\ \ref{Fig:vmaps}) and the 
orphan penumbral area (green contours).  $\sigma$ stands for the 
standard deviation and \# is the number of pixels in each area selected for the study.}
 \label{table:vstatistics}
 \centering
 \begin{tabular}{ccl|ccc|ccc} 
 \hline\hline
     &     &            &    \multicolumn{3}{c}{Spine} & \multicolumn{3}{c}{Orphan penumbrae} \\
 \hline
 Day & Time range & Inversion &       $<v>$         & $\sigma$              & \#       &       $<v>$         & $\sigma$              & \#        \\
     &    (UT)    &      &  (km\,s$^{-1}$)     & (km\,s$^{-1}$)        & points   &  (km\,s$^{-1}$)     & (km\,s$^{-1}$)        & points   \\
  \hline
	 &		  &  Si 		         & $-$ 0.123 & 0.725 & 2097  &     &  & \\
3 Jul.   & 13:53 -- 14:14 &  Si$^\mathrm{mag} (\dagger)$ & $-$ 0.134 & 0.546 & 1974  &  \multicolumn{3}{c}{No orphan penumbrae}   \\
	 &		  &  He 		         & $-$ 0.266 & 0.574 & 146   &        & & \\
			     \hline
	 &		  &  Si 		& $-$ 0.010 & 0.542 & 1114 &  &  &  \\
3 Jul.   & 14:39 -- 15:01 &  Si$^\mathrm{mag}$  & $-$ 0.163 & 0.406 & 1088 & \multicolumn{3}{c}{No orphan penumbrae}  \\
	 &		  &  He 		& $-$ 0.216 & 0.521 & 94   &  &  &  \\
			      \hline
			      \hline
	 &		  &  Si 		& $-$ 0.011 & 0.424 & 1019 & $+$0.086 & 0.383 & 831 \\
5 Jul.   & 7:36 -- 8:05   &  Si$^\mathrm{mag}$  & $-$ 0.190 & 0.333 & 1018 & $-$0.250 & 0.329 & 825  \\
	 &		  &  He 		& $+$ 0.808 & 0.566 & 42   & $+$0.547 & 1.303 & 54 \\
			    	\hline
	 &		  &  Si 		& $+$ 0.108 & 0.326 & 870 & $+$0.026 & 0.353 & 1009 \\
5 Jul.   & 8:08 -- 8:28   &  Si$^\mathrm{mag}$  & $-$ 0.144 & 0.263 & 870 & $-$0.241 & 0.368 & 997 \\
	 &		  &  He 		& $+$ 0.884 & 0.604 & 42  & $+$0.346 & 0.951 & 59  \\
			      \hline
	 &		  &  Si 		& $-$ 0.011 & 0.186 & 780 & $+$0.079 & 0.317 & 1127 \\
5 Jul.   & 8:42 -- 9:01   &  Si$^\mathrm{mag}$  & $-$ 0.237 & 0.288 & 780 & $-$0.243 & 0.353 & 1106 \\
	 &		  &  He 		& $+$ 0.954 & 0.564 & 36  & $+$0.427 & 1.002 & 63\\
			    	\hline
	 &		  &  Si 		& $-$ 0.040 & 0.267 & 760 & $+$0.110 & 0.322 & 1023  \\
5 Jul.   & 9:02 -- 9:21   &  Si$^\mathrm{mag}$  & $-$ 0.291 & 0.246 & 760 & $-$0.229 & 0.336 & 1006 \\
	 &		  &  He 		& $+$ 0.649 & 0.464 & 37  & $+$0.183 & 0.622 & 62\\
			    	\hline
	 &		  &  Si 		& $+$ 0.045 & 0.263 & 569 & $+$0.021 & 0.332 & 1497\\
5 Jul.   & 9:33 -- 9:52   &  Si$^\mathrm{mag}$  & $-$ 0.085 & 0.248 & 569 & $-$0.209 & 0.400 & 1457 \\
	 &		  &  He 		& $+$ 1.013 & 0.645 & 22  & $+$0.366 & 0.826 & 102\\
			    	\hline
	 &		  &  Si 		& $-$ 0.127 & 0.282 & 755 & $+$0.021 & 0.314 & 1292  \\
5 Jul.   & 13:39 -- 13:58 &  Si$^\mathrm{mag}$  & $-$ 0.063 & 0.248 & 755 & $-$0.283 & 0.419 & 1273   \\
	 &		  &  He 		& $+$ 0.779 & 0.960 & 41  & $+$0.278 & 0.814 & 60\\
			    	\hline
	 &		  &  Si 		& $-$ 0.046 & 0.375 & 618 & $+$0.211 & 0.301 & 1257 \\
5 Jul.   & 14:31 --14:51  &  Si$^\mathrm{mag}$  & $+$ 0.014 & 0.290 & 618 & $-$0.099 & 0.427 & 1248  \\
	 &		  &  He 		& $+$ 1.083 & 1.009 & 30  & $+$0.271 & 0.558 & 75\\
  \hline
  \end{tabular}
 \tablefoot{\tablefoottext{\dagger}{Velocities above 2\,km\,s$^{-1}$ or below $-2$\,km\,s$^{-1}$ inferred 
from the ``magnetic'' \ion{Si}{i} inversions (and mostly due to weak $Q$ and $U$ signals)
were excluded from the statistics. However, only very few points 
were affected by this filter ( between 9 -- 40; and 123 for the first map).  }
 } 

 \end{table*}

The inferred line-of-sight velocities of four representative cases (one on July
3 and three on July 5) are shown as slit-reconstructed maps in Fig.\
\ref{Fig:vmaps}. Blue means upflow ($v < 0$) while red indicates
downflow ($v > 0$). The pixels enclosed by the contours have LOS magnetic
field inclinations between $75^\circ \leq \gamma \leq 105^\circ$, 
$\gamma$ being the inclination inferred from the corresponding silicon or helium inversions. 
This criterion was chosen to study the
velocities associated to transverse fields which, as reported in Paper I, dominate along the
filament. The contours denote the area they belong to. The
green contours enclose pixels with transverse field lines which lay at the
orphan penumbral region whereas the black contours correspond to the spine region.
Bear in mind that the FOV on July 5 only overlaps approximately with the
upper half of the July 3 map (the white dashed rectangles in Fig.\ 
\ref{Fig:TIP} give an idea of the common FOV). Moreover, small shifts of the
FOV (in the vertical direction) between different maps of July 5 are also
present. 

The panels in the \textit{upper} row of Fig.\ \ref{Fig:vmaps} correspond to the
\ion{Si}{i} 10827\,\AA\ velocities, scaled between $\pm 1$\,km\,s$^{-1}$,
inferred from the standard inversions, i.e., those with 
$w_{I,Q,U,V} = 1$. We found alternating blue and
red patches in all panels, which revealed a mixture of upflows and downflows along the
PIL, i.e., inside the contours.  However, it is clear that on July 3, the
inside of the black contour is more dominated by upflows than on July 5. 
On this day, there are no significant
differences between the velocity patterns of the orphan penumbrae and the spine. 

We now turn to the panels of the \textit{middle} row of Fig.\ \ref{Fig:vmaps}.
The velocity maps correspond to the magnetic inversions as 
defined in Sect.\ \ref{sect:dataanalysis} ($w_{I,Q,U,V}=0,1,1,0.1$).
We therefore assume that the velocities
in these panels are more likely to represent the motions of the
transverse magnetic field lines (as opposed to the maps from the
first row, which are more influenced by unmagnetized regions and flows along
vertical field lines). The inversion
code had problems in converging properly when Stokes $Q$ and $U$ were small, 
i.e.,  where the magnetic fields were weak or almost
vertical. Nevertheless, along the PIL, the inversion code
performed well and the uncertainties obtained were of the same order of
magnitude as those from the standard inversions. On July 5, a comparison
of the \textit{middle} with the \textit{upper} panels reveals that the
``magnetic'' \ion{Si}{i} velocities inside the contours are globally more
dominated by {\em blueshifts}, rather than redshifts. We interpret this result as
an indication of a predominant upward motion of the transverse field lines 
in the photosphere, in particular inside the orphan penumbrae (green
contours). Similarly, on July 3, 
there is more of an upward trend in the motions of the transverse
fields. 

\begin{figure*}[!t]
\centering
\includegraphics[width=0.7\textwidth]{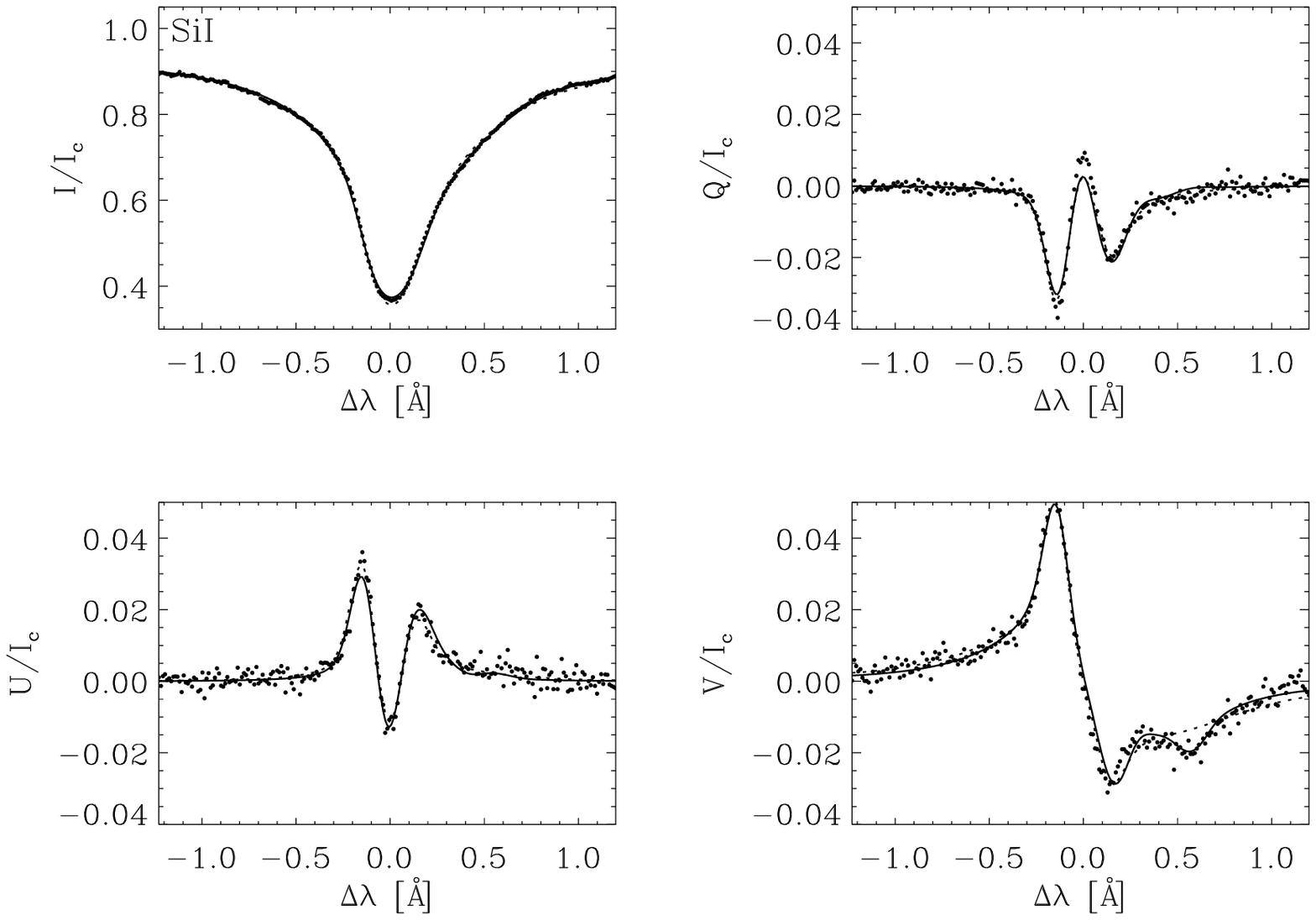}
\caption{From \textit{left} to \textit{right} and \textit{top} to
  \textit{bottom}: Stokes $I$, $Q$, $U$ and $V$ profiles (dots)
 of the \ion{Si}{i} line normalized to the continuum, and best fits (solid line) 
using a two-component SIR inversion. The profiles 
correspond to [$x,y$] = [20\farcs1,19\farcs4] (area 4 in Fig.\ \ref{Fig:maps2comp}). 
The single-component fit (i.e., standard fit) is shown in the dashed 
lines for comparison and clearly does a worse job at fitting Stokes $V$. 
The physical parameters inferred from the two-component inversion are: $v_1 = 0.3 \pm 0.2$\,km\,s$^{-1}$,
$B_1 = 1001 \pm 68$\,G, $\gamma_1 = 67^\circ \pm 3^\circ$, $\phi_1 = 114^\circ \pm 2^\circ$, $f \sim 0.94$ 
for the standard component and $v_2 = 11 \pm 2$\,km\,s$^{-1}$, $B_2 = 1424 \pm 300$\,G,
 $\gamma_2 = 23^\circ \pm 20^\circ$, $\phi_2 = 81^\circ \pm
 51^\circ$, $f \sim 0.06$ for the magnetic one. 
The inferred stray-light is $\alpha \sim 5\%$. 
The uncertainties associated to the parameters of the second component
were obtained with SIR and verified with the sensitivity analysis explained at the end of Section 
\ref{Sect:supersonicdown}.}
\label{Fig:stokes2comp}%
\end{figure*}

The velocity maps inferred from the (binned) \ion{He}{i} 10830\,\AA\ inversions
are presented in the \textit{lower} row of Fig.\ \ref{Fig:vmaps}. Note that the
color is now scaled between $\pm 2$\,km\,s$^{-1}$. The first striking result
are the ubiquitous red areas (indicative of downflows) that extend
almost everywhere, with the exception of the region near the PIL.
On July 3, an almost
perfect correlation between the transverse field lines inferred from the helium
inversions and the upflow areas can be seen within 
the black contour in the \textit{lower lefthand} panel. Bearing in mind that on July 3 we clearly 
saw the filament axis in the helium core absorption map (see Fig.\ 
\ref{Fig:TIP}), and in Paper I we found sheared field lines parallel to its
axis, one can easily deduce that the filament axis is rising in the chromosphere. 
Typical rising speeds are in the range [0,-1.5]\,km\,s$^{-1}$. 

Two days later, on July 5, 
the spine region (black contours) hardly shows any upflows.
The large-scale redshifted pattern seen elsewhere is, however,
weaker. This means that, even if the filament axis was not moving
upwards on July 5, it was still able to interfere with the mechanism that generates
the large scale downflows (see Sect.\ \ref{Sect:coronalrain}).
In the diffuse filament region (above the orphan penumbrae, marked with
green contours), the
velocity pattern is fundamentally different. Once again, the transverse fields
near the PIL harbor clear signs of upflows.
It is thus evident that, while the general trend observed in \ion{He}{i} corresponds
to a large scale downflow, the filament axis (near the PIL) is a place
where the downflows are not as strong and sporadic upflow patches are
also present.
We will discuss this issue later, in Section \ref{Sect:coronalrain}.

To understand the general trends in the velocity maps of Fig.\ \ref{Fig:vmaps}, 
a statistical study of the average velocities ($<v>$) inside the black and 
green contours was carried out and compiled
in Table \ref{table:vstatistics}. The number of averaged points (\#) and 
the standard deviation ($\sigma$) are also provided in this table. 
We have found the following:

\begin{enumerate} 

\item On July 3, the mean velocities inferred from the \ion{Si}{i}
inversions at the spine (i.e., below the filament axis) show an obvious 
blueshift. The magnetic inversions, averaging $<v_\mathrm{Si}^\mathrm{mag}> \sim -0.150$\,km\,s$^{-1}$, 
are more blueshifted than the
standard inversions. Upflows are also found in the
chromosphere ($<v_\mathrm{He}> \sim -0.240 $\,km\,s$^{-1}$).

\item For the first four maps of July 5, the spine presents an upward
motion of the photospheric transverse fields (black contours) 
as inferred from the magnetic inversions, with typical velocities of
$<v_\mathrm{Si}^\mathrm{mag}> = -0.215$\,km\,s$^{-1}$. In the last
three maps the velocities drop to almost zero.  
The standard \ion{Si}{i} inversions show, by contrast,
velocities which oscillate around zero
($<v_\mathrm{Si}> \sim -0.012$\,km\,s$^{-1}$). 

\item On July 5, the chromospheric velocities at the spine are
dominated by downflows in the range of
$v_\mathrm{He} \in [0.65,1.08]$\,km\,s$^{-1}$. We note, however,
that these downflows are smaller than the typical values observed
elsewhere in the surrounding facular region (see Sect.\ \ref{Sect:coronalrain}).

\item At the orphan penumbrae (green contours), there is a
systematic difference between the photospheric LOS velocities
inferred from the ``standard'' and those obtained from the 
``magnetic'' inversions of the
\ion{Si}{i} line. While the latter clearly
show upflows in all maps, with values between $v_\mathrm{Si}^\mathrm{mag} \in
[-0.099, -0.283]$\,km\,s$^{-1}$ and an average of $<v_\mathrm{Si}^\mathrm{mag}> =
-0.222$\,km\,s$^{-1}$, the standard inversions show slightly slower velocities,
with downward motions in the range of $v_\mathrm{Si} \in [0.021,
0.211]$\,km\,s$^{-1}$, and a mean value of $<v_\mathrm{Si}> =
0.079$\,km\,s$^{-1}$. 

\item Above the orphan penumbrae (inside the green contours) the average
velocities inferred from the \ion{He}{i} inversions show
downflows in the range of $v_\mathrm{He} \in
[0.18,0.55]$\,km\,s$^{-1}$. These downflows are weaker
than those observed in the spine (see point 3 above).

\end{enumerate}

\subsection{Two-component Stokes $V$ profiles at the PIL: supersonic downflows} \label{Sect:supersonicdown}
\begin{figure*}[!t]
\centering
\includegraphics[width=0.6\textwidth]{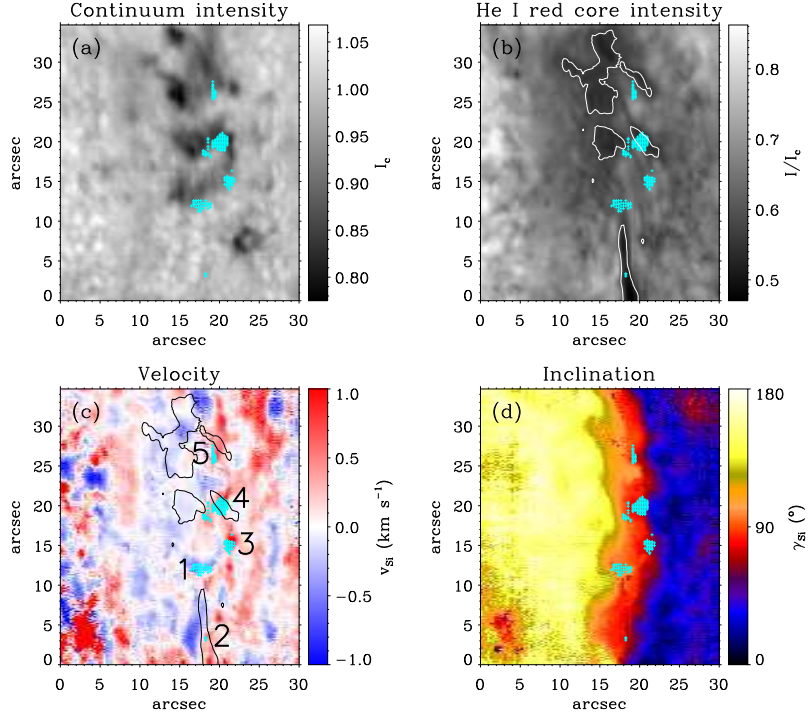}
\caption{(a) The continuum intensity image show pores and orphan penumbrae at the PIL. (b) The absorption in helium 
(red core) clearly shows the spine in the lower part of the image and a diffuse filament in the upper part. (c) LOS
velocity map inferred from the standard single-component inversions of SIR. 
The marked areas (1--5) correspond to the 
two-component Stokes $V$ profiles (the average inclinations and velocities for these areas are presented in 
Table \ref{table:2comp}). (d) LOS inclinations from the single-component inversions. 
In all maps, the cyan colored crosses mark the positions where two-component profiles were detected.  }
\label{Fig:maps2comp}%
\end{figure*}

On July 5, the day when we first observed the orphan penumbral
system, we detected the presence of atypical multilobed Stokes $V$ profiles (e.g., Fig.\ 
\ref{Fig:stokes2comp}) in the \ion{Si}{i} 10827\AA\ line. These multilobed
profiles are a clear indication of the presence of several magnetized
components within the resolution element. Most of these multi-component
profiles were found near the borders of the orphan penumbrae. To illustrate
this, we selected one map from July 5 (taken between 13:39--13:58 UT) and
marked the location of all two-component profiles with small cyan crosses (see Fig.\ \ref{Fig:maps2comp}). 
The other maps from July 5
show a similar spatial distribution of the multi-component
profiles. Only points with a Stokes
$V/I_\mathrm{c}$ signal of at least $0.01$ were selected for this
study. In this way, we made sure that the origin of these profiles was
related to a multiple-velocity component scenario, and we were able to
rule out other causes such as mixed polarities, which are also a very common
occurrence in the proximity of PILs.
Figure \ref{Fig:maps2comp}(a)
shows a continuum intensity image. The two-component profiles are located
on top of pores and orphan penumbrae, between $x \in [16\arcsec,22\arcsec]$.
The helium red component intensity map in Fig.\ \ref{Fig:maps2comp}(b)
shows that the largest group of atypical Stokes $V$ profiles is
co-spatial with the dark
\ion{He}{i} thread located around $[x,y] = [20\arcsec,20\arcsec]$. Only
three asymmetric profiles were detected at the spine, around $[x,y] =
[18\farcs5,3\arcsec]$.  The contours in Figs.\  \ref{Fig:maps2comp}(b-c)
correspond to the darkest areas, i.e., highest absorption of the \ion{He}{i}
red component, of Fig.\  \ref{Fig:maps2comp}(b).  Bear in mind that the
multi-component profiles were only found in the photosphere; there is no
signature of these highly Doppler shifted components in the chromospheric
\ion{He}{i} 10830\,\AA\ profiles. The atypical \ion{Si}{i} profiles contain
information about a strongly redshifted component that the standard SIR
inversions cannot pick out. The conspicuous signature present in Stokes $V$
calls for a two-atmosphere inversion that can provide some insights into
the nature of this shifted component. Such inversions were performed on those
specific profiles where this signature was visually detected. The following
properties were inferred:

\begin{enumerate}

\item The atypical Stokes $V$ profiles appear at, or very close to, the
PIL, i.e., where the magnetic field lines are mainly
horizontal (see inclinations inferred from the standard single-component
silicon inversions, $\gamma_\mathrm{Si}$ in Fig.\ \ref{Fig:maps2comp}(d)). 
 
\item From the two-component SIR inversions we inferred: (1) a 
dominant transverse component with a filling factor in the range of $f_1 \in [0.87,0.98]$
and inclinations with respect to the LOS  $\gamma_1 \in
[68^\circ, 95^\circ]$. This is the component that is most similar to that retrieved
by the standard inversions.
(2) A second component with smaller filling factors, $f_2 \in [0.02,0.13]$,
oriented more parallel to the LOS, with inclinations  
$\gamma_2 \in [21^\circ, 58^\circ]$. 

\item The inferred Doppler maps show that the dominant transverse component 
has velocities in the range of $v_1 \in [-0.1,0.5]$ km\,s$^{-1}$ while the
longitudinal one, with smaller filling factors, has downward velocities 
$v_2 \in [6.2,11.9]$ km\,s$^{-1}$. It is clearly a supersonic downflowing
component.

\item In Fig.\ \ref{Fig:maps2comp}, all the detected atypical Stokes $V$
profiles had a second component with the same polarity than the first one,
except for the group of points between $y = [11\arcsec,13\arcsec]$ (area 1 in
Fig.\  \ref{Fig:maps2comp}(c)) which had a second component with opposite
polarity to that of the first one. Note, however, that the dominant
polarity, being highly transverse, is very much prone to polarity changes
due to projection effects.

\end{enumerate}

 \begin{table}[!t]
 \caption{Mean LOS inclinations ($<\gamma>$) and velocities ($<v>$) inferred from the two-component 
inversions of the
\ion{Si}{i} line. The different areas are marked in Fig.\ \ref{Fig:maps2comp}(c). $\sigma$ stands for the 
standard deviation and \# is the number of pixels averaged in each area.}              
 \label{table:2comp}      
  \centering
 \begin{tabular}{ccccccc} 
 \hline\hline
 Area & Comp. & $<\gamma>$  & $\sigma_\gamma$ & $<v>$          & $\sigma_v$     & \#     \\
      &           & ($^\circ$)  & ($^\circ$)      & (km\,s$^{-1}$) & (km\,s$^{-1}$) &   (px) \\
  \hline
 \multirow{2}{*}{1} & 1 & 94.6 &  1.4                   &  $-$0.126 & 0.135 & \multirow{2}{*}{25}     \\
                    & 2 & 57.7 &  51.5                  &  11.890 & 8.867 &      \\
\hline
 \multirow{2}{*}{2} & 1 & 76.3 & 2.3                    & 0.351  & 0.034 &   \multirow{2}{*}{3}   \\
                    & 2 & 20.7 & 20.2                   & 4.699  & 2.439 &      \\
\hline
 \multirow{2}{*}{3} & 1 & 68.5 & 6.5                    & 0.496  & 0.315 &   \multirow{2}{*}{32}   \\
                    & 2 & 43.7 & 18.1                   & 7.595  & 2.278 &      \\
\hline
 \multirow{2}{*}{4} & 1 & 72.7 & 11.8                   & 0.355  & 0.256 &   \multirow{2}{*}{85}   \\
                    & 2 & 28.4 &  17.4                      & 9.632  & 2.795 &      \\
\hline
 \multirow{2}{*}{5} & 1 & 78.7 & 3.5                    & 0.250  & 0.113 &  \multirow{2}{*}{18}    \\
                    & 2 & 49.4 & 20.4                   & 6.190  & 2.919 &      \\
\hline
  \end{tabular}
 \end{table}

\begin{figure*}[!t]
\centering
\includegraphics[width=0.8\textwidth]{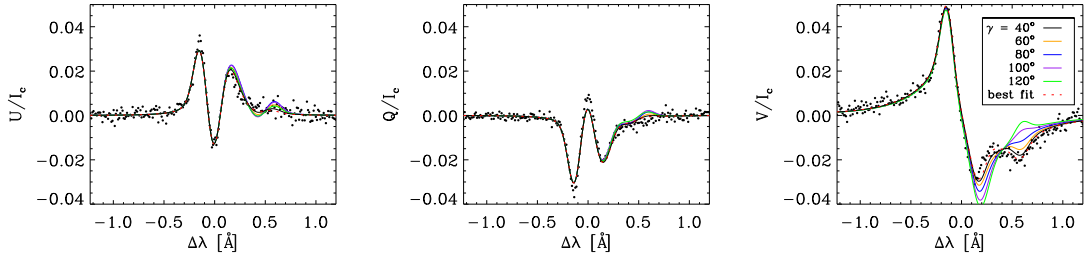}
\caption{Dots represent the observed \ion{Si}{i} Stokes $U$, $Q$, and $V$ profiles (same as Fig.\ \ref{Fig:stokes2comp}). 
The overplotted synthetic profiles were computed using different
magnetic field inclination angles for the second component, 
that varied between $\gamma \in$ [40$^\circ$, 120$^\circ$]. 
The test was performed by varying the magnetic
field inclination of the second component and keeping the remaining
atmospheric parameters (which came 
from the best fit of the SIR inversion) fixed. 
The dotted red line shows the best matching profile, whose
corresponding atmospheric parameters are shown in the caption
of Fig.\ \ref{Fig:stokes2comp}.}
\label{Fig:comptests}%
\end{figure*}

The multiple component profiles were grouped in five separate
areas. Since the properties of the dominant and of the redshifted components
differ in each of these areas, a separate statistical study for them has
been carried out.
The five areas are marked in Fig.\ \ref{Fig:maps2comp}(c).  Table \ref{table:2comp}
summarizes the mean LOS inclinations ($<\gamma>$) and velocities ($<v>$), for both
components, in the five areas. The standard deviation ($\sigma$)
of both quantities, as well as the number of pixels used for the statistics (\#),
are also listed. As stated before, the second component
is completely dominated by magnetic field line inclinations which are more
longitudinally orientated. These inclinations, however, vary substantially
within each area (note the large values of the standard deviation
$\sigma_\gamma$ for Comp.\ 2 in Table \ref{table:2comp}).  
In these atypical cases, the inversion code provides considerable
uncertainties associated with the atmospheric parameters, 
something that should be expected since this redshifted component has a small
effect on the observed profiles, as reflected by its low filling
factor (mostly $< 0.1$). However, a clear tendency for smaller inclinations, i.e., 
more vertically oriented field lines, is seen in all five areas.

  \begin{figure*}[!t]
 \centering
   \includegraphics[width=0.8\textwidth]{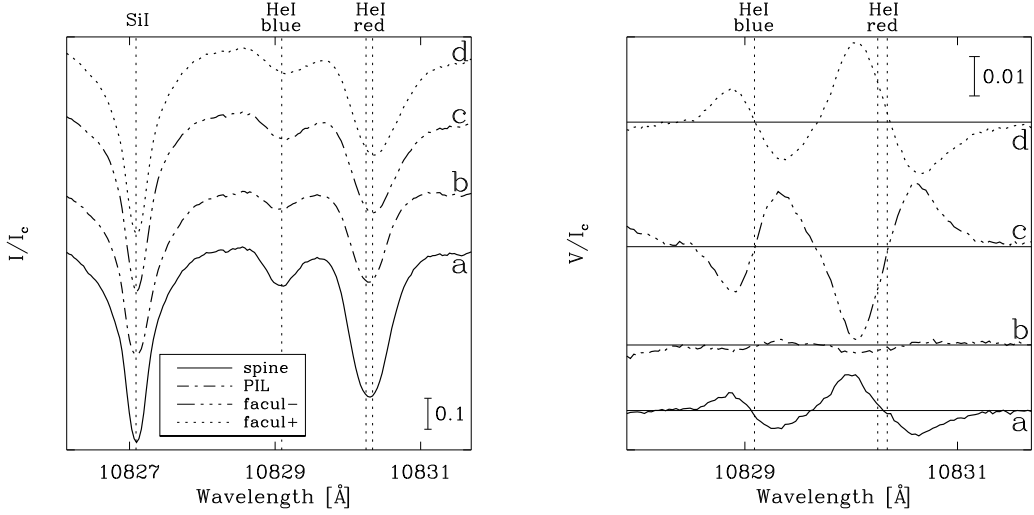}
    \caption{\textit{Left}: averaged intensity profiles $I/I_\mathrm{c}$ of four selected positions 
of the map acquired on July 5 between 13:39-13:58\,UT. Each profile is marked with a character (a--d) which indicates
  its position in Fig.\ \ref{Fig:TIP}. Note the shifted red wing of the blended
 \ion{He}{i} red component for the two facular
  profiles indicating the presence of more than one component contributing to the line formation. \textit{Right:}
 the corresponding helium Stokes $V/I_c$ profiles to the intensity profiles shown on the \textit{left} panel. 
The solid lines along the $x$-axis indicate  $V/I_c = 0$ for each profile. 
The dotted vertical lines, in both panels, mark the position of the line 
cores of the involved ions.  }
    \label{Fig:coronalrain}
    \end{figure*}

Given the small amplitude of the signatures that this component leaves on the emergent Stokes profiles,
the question is raised of how reliable the inversions are. To address
this question, we performed the following test:
We selected several pixels with multilobed Stokes $V$ signals
and used the atmospheres resulting from the two-component inversions
to synthesize sets of Stokes profiles. 
For each model atmosphere, several different realizations of the synthesis, with different
values of the inclinations and azimuths of the second component, were carried out
(the first component was kept unchanged).
Figure \ref{Fig:comptests} shows one example of the
synthetic Stokes $Q$, $U$, and $V$ profiles (Stokes $I$ showed no changes and
was therefore not presented) with different inclinations, between 40$^\circ$
and 120$^\circ$. Stokes $Q$ and $U$ present minor changes in
their shape that are just above the noise level and
therefore not conclusive. However, the changes in inclination substantially
affected the resulting Stokes $V$ profiles. It is apparent from Fig.\ \ref{Fig:comptests}
that the most likely inclination is far from being transverse. In
fact, the best fit (dotted red line) yields an inclination
of $\gamma = 22.9^\circ \pm 20^\circ$. We also carried out the same test
changing the azimuth. Stokes $Q$ and $U$ were only slightly
affected by these variations, again within the noise level, while Stokes $V$
was completely insensitive to the changes. This shows that while the inclination
of the redshifted component is well established (and that it is more vertical
than the dominant, transverse, component), the value of its azimuth is 
highly uncertain.

\subsection{Ubiquitous downflows around the PIL in the helium velocity maps} \label{Sect:coronalrain}

As mentioned above, the velocities inferred from the \ion{He}{i} inversions
outside the PIL (in the faculae) are ubiquitously dominated by downflows on both
days (see the lower panels of Fig.\ \ref{Fig:vmaps}). This
was largely unexpected, so it is important
to understand what signature in the spectral profiles made MELANIE infer
these downflows. Figure \ref{Fig:coronalrain} shows four
representative Stokes $I$ and $V$ profiles from different areas of the map taken
between 13:39--13:58\,UT on July 5. They correspond to: (a) the spine,
(b) the PIL, (c) the negative and (d) the positive polarities within the faculae. 
The locations of pixels a--d are marked in Fig.\ \ref{Fig:TIP}. The dotted
vertical lines, in Fig.\ \ref{Fig:coronalrain}, mark the centers of the line
cores at rest. In the \textit{lefthand} panel, the intensity spectra comprise the
photospheric \ion{Si}{i} line and the \ion{He}{i} triplet. The cores of the
four silicon intensity profiles are almost perfectly centered at their
rest wavelengths. However, while the \ion{He}{i} triplet profiles of the
spine and the PIL are also approximately at rest, the two facular profiles 
are clearly redshifted. 
It is also of interest that the red wing of the \ion{He}{i} red component of the
facular profiles is stretched away from the line core. This, together with the
fact that the profiles are not symmetric, indicates that more than one atmospheric 
component was involved in the line formation
process, and that one of these components
harbors distinct downflows.  The redshift in the
helium red component of the facular regions is also seen in their Stokes $V$
profiles (c and d) in the \textit{righthand} panel of Fig.\ \ref{Fig:coronalrain}.
As expected, profile (b) has almost no Stokes $V$ signal since it is located 
at the PIL, where the transverse fields dominate.   

It is, thus, clear that the facular profiles seen in the chromosphere are
substantially redshifted. A possible explanation for this is given
in Sect. \ref{Sect:Discussion}. To quantify the general trend of 
the facular velocities found in our data sets we did averages only taking into
account those pixels which had photospheric inclinations close to longitudinal ($\gamma_\mathrm{Si} < 25^\circ$ and
$\gamma_\mathrm{Si} > 155^\circ$). Note that the inclinations we used are the
ones inferred from the \ion{Si}{i} 10827\,\AA\ inversions, whereas the
velocities correspond to the ones obtained from the \ion{He}{i} 10830\,\AA\
inversions. This criterion ideally distinguishes the points belonging
to the faculae \citep[having mainly longitudinal fields at photospheric heights,]
[]{martinez97} from those of other areas, such
as the PIL or outside the faculae. The mean velocities found in the faculae,
for all maps (July 3 and 5), were in the range of
$v_\mathrm{He}^\mathrm{fac} \sim 1.4 - 1.8$\,km\,s$^{-1}$. The
standard deviations associated to these velocities were typically around $\sigma =
1.0 - 1.4$\,km\,s$^{-1}$, and the number of points used for the
statistics of each map was between $300 - 400$. As mentioned above, since these \ion{He}{i}
intensity profiles are likely to comprise more than one component,
multi-component inversions in this spectral region
\citep[e.g.,][]{lagg07,sasso11} might reveal slightly stronger downflows in the
faculae.

\subsection{Time series}

\begin{figure}[!h]
   \resizebox{\hsize}{!}{\includegraphics{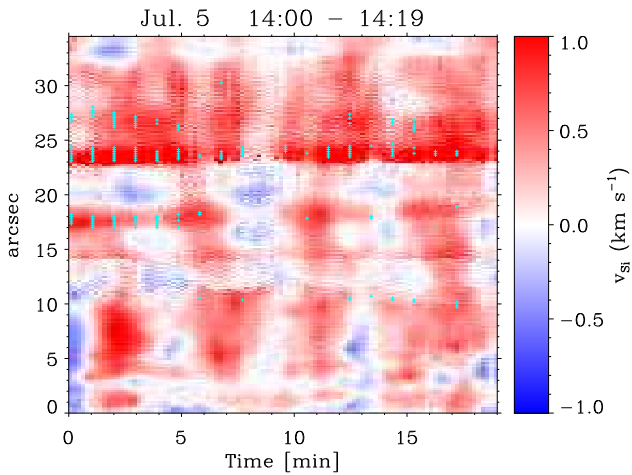}}
    \caption{Photospheric velocity time evolution. The time series was taken with the slit fixed on top of the PIL
 (between $x \sim 18\arcsec - 20\arcsec$ in the second column of Fig.\ \ref{Fig:TIP} ). The velocities were inferred 
  from the ``standard'' \ion{Si}{i} 10827\,\AA\ inversions.
  The cyan-colored crosses mark the positions of the two-component profiles. }
    \label{Fig:timeseries}
    \end{figure}

A time series with the slit fixed over the PIL was acquired on July 5,
approximately between $x \sim 18\arcsec - 20\arcsec$ on the \textit{righthand}
panel of Fig.\ \ref{Fig:TIP}. The velocities shown in Fig.\ \ref{Fig:timeseries} were
retrieved from the standard \ion{Si}{i} inversions.
Time is represented along the $x$-axis while
the $y$-axis shows the spatial direction along the slit. In the lower half of
the figure, the photospheric 5 minute oscillation is unequivocal.
This corresponds to the area below the spine. There is an apparent
phase shift of the oscillation pattern, starting at $y \sim 15\arcsec$.
This shift coincides with the beginning of the orphan penumbrae. The single
most striking finding from this figure was the continuous downflow
seen between $y \sim 23\arcsec - 25\arcsec$ (dark red pattern) throughout the whole
time range, with which not even the photospheric oscillation
pattern interfered. 

The Stokes $V$
profiles were carefully inspected one by one along the slit every minute. This was done
in order to find the atypical multilobed Stokes $V$ profiles defined in Sect. \ref{Sect:supersonicdown}.
As in Fig.\ \ref{Fig:maps2comp}, the cyan crosses in Fig.\ \ref{Fig:timeseries} represent the
location of the detected atypical Stokes $V$ profiles.
From this figure it is obvious that the lifetime of the
supersonic downflow was at least as long as our time series (19 minutes).  
We take this time to be a
lower limit but it might, in fact, be much longer, since this supersonic
downflow was not an isolated event. Other strong downflows have been
detected on all maps from July 5 (albeit not continuously), starting from 7:36
until 14:51\,UT.

\section{Discussion}\label{Sect:Discussion}
In this paper, we have investigated the line-of-sight velocities inferred from
chromospheric and photospheric spectropolarimetric inversions of the
Stokes profiles measured in and below an
active region filament. The aim of this study, together
with the results of Paper I, is to clarify the formation and evolution of this
AR filament. In \mbox{Paper I} we reported that the magnetic field
configuration of the spine of the filament is compatible with a flux
rope topology (on July 3, and also in the spine portion of July 5).
Furthermore, the orphan penumbral region of July 5, was
also likely to be the imprint of a flux rope, but with its axis sitting
low in the photosphere.
These results are supported by recent magnetic field
extrapolations of the same filament by \citet{yelles12} who, for July 5, 
also deduced the presence of a flux rope structure whose axis
lay below the formation height of the \ion{He}{i}
10830\,\AA\ lines. Our AR filament is, therefore, extremely low-lying.

The next question that needs to be addressed is the mechanism by which flux
ropes in AR filaments are formed.  First of all, at the time of the observing
run, the filament had already reached the chromosphere. However, on July 5,
the low-lying magnetic structure and the newly appeared pores and orphan
penumbrae indicated that new flux had emerged between the two days of observation. This flux was mainly
horizontal due to its proximity to the polarity inversion line, as reported in Paper I.
Thus, we placed special emphasis on the study of the
photospheric LOS velocities associated with these transverse fields, i.e., 
inferred from what we called the silicon ``magnetic'' inversions, which provided
information similar to that of the Dopplergrams of the magnetic component presented by
\citet{okamoto09} and \citet{lites10}. There is a very good agreement
among the velocities from all maps of July 5 (see Table
\ref{table:vstatistics}).
All the maps but one pointed to an upward movement of the photospheric transverse
fields, with average velocities between $<v_\mathrm{Si}^\mathrm{mag}> = -
0.21$ and  $- 0.28$\,km\,s$^{-1}$, the last map being the exception, where slower
upward motions ($\sim -0.10$\,km\,s$^{-1}$) are shown. These velocities agree surprisingly
well with the ones presented by \citet{okamoto09} ($-0.30 \pm
0.20$\,km\,s$^{-1}$). However, when carrying out ``standard'' inversions,
(weighting all four Stokes parameters in the
inversion equally) the
average inferred velocities were close to zero or
positive (downflows), with values between $<v_\mathrm{Si}> = 0.02 -
0.11$\,km\,s$^{-1}$ and $0.21$\,km\,s$^{-1}$ for the last map. 
Note that the differences between both inversions
is, on average, of around 0.3\,km\,s$^{-1}$. This is indeed a small
number. It corresponds to 11 m\AA\ or, equivalently, a sixth of the spectral resolution. The fact 
that we obtain velocities that are more redshifted when 
we weight Stokes $I$ and $V$ equally in the inversion
indicates that a longitudinal
component harboring downflows is present, to some degree, in all pixels.
While no evidence for this is provided here, we hypothesize that the
same chromospheric downflow that is observed almost everywhere in \ion{He}{i}
has an impact on the \ion{Si}{i} data, producing slightly more redshifted
values when all Stokes parameters are equally weighted in the inversion code.

In the chromosphere above the orphan penumbrae, the retrieved mean LOS
velocities are dominated by 
downflows in the range of $<v_\mathrm{He}>
= 0.18 - 0.55$\,km\,s$^{-1}$. Nevertheless, these averaged velocities need to
be interpreted with caution since the helium velocity maps of Fig.\ 
\ref{Fig:vmaps} clearly show localized upflows and slower downward motions at
the PIL (inside the green contours). These upward motions at the PIL reach
velocities of about $- 1$\,km\,s$^{-1}$ in all the maps of July 5. A
continuous upflow of mass, lasting at least $\sim 7$ hours, is detected at some
locations of the PIL.  Since helium traces the upper part of the flux rope
(whose axis is lying below, in the photosphere), these localized upflows strongly
suggest that the material is being pushed upwards by the magnetic field lines of the
top of the flux rope \citep[similar to what is seen in the simulations
by][]{sykora08}.  This is in agreement with previous simulations of flux rope
emergence \citep[e.g.,][]{fan01,manchester04,archontis04,fan09}, 
where only the upper field lines are able to rise and expand in
the corona, leaving the main body of the flux tube behind. This
process is driven by the so-called Parker instability.
This interpretation would lead to plasma drainage wherever
the field lines are not horizontal or dipped, and, hence, would explain the
photospheric downflows close to the PIL. 
Sometimes, the downflows at the orphan penumbrae
produce atypical multilobed Stokes $V$ profiles. Examples of these were
found in some areas near the PIL (see Fig.\ \ref{Fig:maps2comp}). Two-component
inversions of these Stokes $V$ profiles yielded mainly
supersonic speeds, in the range of $6 - 12$\,km\,s$^{-1}$, 
for the redshifted component. Similar strong downflows have
been detected in the past
\citep[e.g.,][]{valentin94,lites02,shimizu08} and
have also been found in numerical models of emerging flux regions \citep{cheung08}.
The case described by \citet{valentin94} is particularly
relevant to this paper because it describes a downflow below an AR filament,
similar to the ones studied here. Magnetic
reconnection is one of the proposed mechanisms to cause these supersonic
downflows as suggested from the simulations of \citet{cheung08}. 
In our data sets, this could only have 
taken place in the photosphere which is where we see them clearly. 
The long periods of time (tens of minutes)
during which the redshifted flows persist basically unaltered, almost rule out any 
link with a fast process such as reconnection. The downflows might also
be interpreted as a combination of plasma draining from the flux
rope's body during its emergence process at the photosphere
and/or the continuation of the ubiquitous chromospheric downflows. 

The filament formation model of \citet{vanballe89} 
would expect horizontal field lines submerging at the PIL after a
magnetic reconnection event \citep{vanballe07}.  
While strong downflows in our observations are clearly
detected at several positions along the PIL, they are never co-spatial
with transverse magnetic fields. The inclinations retrieved from the
inversions of the atypical Stokes $V$ profiles,
showed that the second component (the strongly redshifted one) was much more
longitudinally orientated ($\gamma_2 \in [21^\circ,58^\circ]$) than the first one
($\gamma_1 \in [68^\circ,95^\circ]$) (see more details in Table
\ref{table:2comp}). Our test inversions support the reliability of these
results (see Fig.\ \ref{Fig:comptests}). Therefore,
we can safely say that our AR filament presents no evidence of
submerging horizontal field lines, thus, challenging some features of
the aforementioned models.

Let us consider now the spine of the filament, which was seen in
the middle of the FOV on July 3 and only in its
lower part on July 5. This portion of the filament is what we
would call a classical case of AR filament, ``classical'' meaning that it has a
thin elongated shape presenting a flux rope topology and is located
in the chromosphere (it is perfectly discernible in the \ion{He}{i}
absorption image of Fig.\ \ref{Fig:TIP}).
There is a distinct difference between the chromospheric LOS velocities
of both days. On July 3, the spine is completely dominated by upflows
which perfectly trace the transverse fields (see black contour in the
\textit{lower left} panel of Fig.\ \ref{Fig:vmaps}). The average upward velocity
inside this black contour was $<v_\mathrm{He}> \sim - 0.24$\,km\,s$^{-1}$. 
On July 5, however, the spine is not as
strongly blueshifted, characteristic that continues throughout the day (see Fig.\ 
\ref{Fig:vmaps}). Downflows in the range of $<v_\mathrm{He}> \sim 0.81$ and
$1.09$\,km\,s$^{-1}$, are detected in the $\sim 7$\,hours between the
first and last maps.
We interpret this result as a halt in the rise of the filament's
axis. This event makes the spine area be dominated by the same
ubiquitous chromospheric downflows that are seen in the rest of the FOV.

The photospheric motions of the spine on July 3 
show the same trend as the chromospheric ones, with
average upflows of the transverse fields of $<v_\mathrm{Si}^\mathrm{mag}> \sim - 0.15$\,km\,s$^{-1}$.
We interpret this as a coordinated upflow of the entire flux rope. 
This is compatible with the flux rope emergence simulations,
in which the whole structure is very low-lying.
On July 5, the transverse fields begin the day moving upwards $<v_\mathrm{Si}^\mathrm{mag}>
= - 0.19$\,km\,s$^{-1}$, but after $\sim 2$\,hours they seem to stop,  still remaining
at rest by the end of the observing run, four hours later. 

We found ubiquitous downflows in the chromosphere, on both sides of the
polarity inversion line. This was consistent throughout all of our observations. 
The average line-of-sight velocity in the faculae, for all maps, was
$1.6$\,km\,s$^{-1}$ with a dispersion of $\sim 1.2$\,km\,s$^{-1}$. 
We hypothesize that these downflows are a manifestation of 
the so-called \textit{coronal rain}.
This phenomenon occurs when plasma condenses in the corona and then
flows along coronal loops, down into active regions \citep{tandberg95}. So far,
however, our knowledge of this process is still rather poor 
\citep[but see][]{antolinR11}. Recent studies
reveal that the velocities found in coronal rain are in a range between
$20-120$\,km\,s$^{-1}$ \citep{antolin10, antolin11,antolinR11}. This is much faster than
the velocities obtained from our data (Fig.\ \ref{Fig:vmaps}), although
our measurements correspond to lower heights. There is a strong
possibility that more than one atmospheric component, along the line-of-sight,
contributes to the \ion{He}{i} 10830\,\AA\ triplet formation. This can be
deduced from the shape of the red wings of the intensity
profiles in the faculae (\textit{lefthand} panel; Fig.\ \ref{Fig:coronalrain}).  However, the
signature is weak and, hence, it is unclear whether 
strong downflows, such as those reported by
previous authors in the \ion{He}{i} triplet \citep[$\sim 42$\,km\,s$^{-1}$,
e.g.,][]{muglach97, schmidt00, lagg07}, would be inferred
in a multi-component analysis. Another 
explanation could be that material is falling from the slowly rising flux rope structure on
July 3, or from the expanding field lines above the orphan
penumbrae on July 5, following vertical magnetic field lines 
\citep[in a similar way as the draining of
rising loops proposed by][]{lagg07}. However, these processes can also
be considered to be different manifestations of the coronal-rain phenomenon.

\section{Conclusions}
For the sake of clarity we divided the study of the AR filament into two parts,
the main distinction between them being the height at which
the filament axis lies. The first part corresponds to the spine
(seen mainly on July 3, and also in the lower half of the FOV on July 5), where the
filament axis lies in the chromosphere. The second corresponds to the diffuse 
filament (seen only in the upper half of the FOV on July 5), which sits above the
orphan penumbrae and pore regions, and has a much lower-lying filament axis.
In this latter case, the helium 10830\,\AA\
only traces the upper part of the flux rope, as explained in 
Paper I. The main conclusions of this study are:

\begin{enumerate} 
\item On July 3, the LOS velocities inferred from
the helium and ``magnetic'' silicon inversions (that trace the
behavior of the horizontal magnetic fields) in the \textit{spine}
region, show generalized upflows, which we interpret
to represent the emergence of the flux rope structure as a whole.   
Two days later, on July 5, the spine shows downflows in the chromosphere, similar
to those seen elsewhere in the facular region. Yet the photospheric
velocities in this region present upflows for the first few hours of the
day that drop to velocities close to zero towards the end of the
observing run.


\item On July 5, in the chromosphere, above the orphan penumbrae, the 
LOS velocities are on average redshifted. This redshift is, however,
smaller than the dominant redshift seen in the facular region. Indeed,
blueshifted patches are present along the PIL in all data sets. We propose
that the blueshifted patches seen in the chromosphere are due to field lines
which expand from the lower-lying flux rope into the chromosphere, similar to
what is found in the flux rope emergence simulations \citep[e.g.,][]{sykora08}.
The photospheric transverse field lines along the orphan penumbrae are clearly
moving upwards. This is consistent in all seven data sets. Therefore the
sheared field lines (the flux rope axis) are rising.  

\item Atypical multilobed Stokes $V$ profiles were found in the photosphere near the PIL.  
Two-component inversions of these profiles revealed
localized supersonic downflows in the strongly redshifted component. These
downflows last for at least 19 minutes, ruling out any episodic origin. 
Furthermore, the retrieved magnetic fields associated with this component are
oriented along (or very close to) the line of sight.
Therefore, we cannot identify this
component with submerging loops that harbor horizontal fields as
in the models from \citet{vanballe89} and \citet{vanballe07}.  
It is also important to point out that the number of detected two-component profiles 
was only a small percentage of the total number of points inside the PIL (12 \%), which, on
average, showed an upward motion of the transverse fields.    

\item Almost ubiquitous redshifts of the chromospheric \ion{He}{i} 10830\,\AA\,
triplet, with average downflows of $\sim 1.6$\,km\,s$^{-1}$ and a dispersion of $\sim
1.2$\,km\,s$^{-1}$, were found in the faculae in all data sets.

\end{enumerate}

The global picture resulting from the Doppler shifts studied in
this paper can be summarized as follows. AR faculae are immersed 
in a global rainfall of mass from the upper layers, 
as indicated by the redshifted \ion{He}{i} line. We have tentatively associated
this redshift with coronal rain. This process has no evident counterpart in the
photosphere, being some localized supersonic downflows the only possible candidate. 
In this environment, the transverse field lines
(including the filament axis observed either in the chromosphere or in the 
photosphere) display upflows. The only case where this does not
hold true is that of the chromospheric spine region on 
July 5. In this area, only the global downflow is observed.

As we have
learned from the simulations, the process of flux rope emergence through the 
photosphere is not an easy one and will often be aborted.
We propose that this is what might have happened on July 5 at the spine region.
Note, however, that the rest of the filament for this day displayed clear
upflows at both heights. All in all,
the present study supports the scenario of an emerging flux rope from below the
photosphere. Vector magnetograms, as well as LOS velocities in the photosphere
and chromosphere, agree with the proposed scenario. Whether or not this emergence
process is common to all active region filaments, or at least to those that
present a photospheric manifestation (orphan penumbrae), needs to be
proven by using further multi-wavelength and multiheight observations. 

There are some limitations in this work that need to be considered. First, an important 
observational gap on July 4 exists. Data
from that day would have contributed to a better understanding of the
inferred magnetic structure observed on July 5. Another limitation to this
work is the small field of view of the polarimeter that we had at the
time of the observations (nowadays the slit is twice as large). 
This work has proved that there
is an imperative need for multi-wavelength instruments and
larger FOVs in order to understand the formation process and evolution
of AR filaments. It is crucial to have
magnetic field information of at least two heights, e.g., in the photosphere
and in the chromosphere, to carry out a simultaneous and co-spatial
analysis of the evolution of the filament. 

The flux rope that constitutes the studied AR filament is extremely
low-lying, especially the portion observed on the second day.  
This result makes it hard 
for the filament formation models that build flux ropes in the corona
by reconnection \citep[e.g.,][]{vanballe89,DeVore00} to
reproduce the scenario that we observe.
In particular, the large scale submergence of photospheric transverse field
lines is not observed in our data. 
While a rather complete picture of the evolution of this AR filament
(favoring a flux rope emergence from below the photosphere) has come out of
this series of papers, we would like to stress that the formation process
of AR filaments in general could differ, maybe substantially, from this case.

\appendix
\section{Velocity calibration} \label{app:a}

In order to obtain LOS velocities on an absolute scale, our 
Doppler shifts have to be calibrated to high precision and
corrected for the systematic effects introduced by the Earth's rotation, the orbital motion of Earth around the
Sun, the solar rotation, as well as for the solar gravity redshift. 

The final accuracy and precision have to be high enough to be
able to measure typical photospheric Doppler shifts which, in the
case of the \ion{Si}{i} line, correspond to velocities in the range of few hundred m\,s$^{-1}$.
Since our spectral range comprises two telluric H$_2$O lines, we can use these
to calculate our sampling (\AA\ per pixel). A Gaussian function with six terms was
fit to the deepest part of these lines in our data in order to
determine the position of their centers.
The difference, $\Delta x$, is the distance between both lines in pixel units. 
The same process was carried out on the Fourier Transform Spectrometer (FTS) spectrum
\citep{FTS} from the Kitt Peak National
Observatory, obtaining $\Delta \lambda = 1.873$\,\AA\ for the distance between both lines. 
The spectral sampling is merely calculated by dividing $\Delta \lambda$/$\Delta x$.  
This was done for each map using the average spectrum from a small 
non-magnetic region (containing about 1000 pixels). 
The mean spatial sampling of all maps is $11.035 \pm 0.010$
m\AA/px. Ideally, the sampling should always be the same. It can be
theoretically calculated using the specifications of the telescope and of the
spectrograph. A comparison with the theoretical sampling ($11.240$ m\AA/px)
reveals that the calibration delivered a slightly smaller value,
but not too different.

We are now able to construct the wavelength array using the calculated
spectral sampling and a telluric line whose central wavelength is known, 
as a reference. According to the solar spectrum atlas of \citet{spectrum70}, the telluric line closest to the
\ion{He}{i} triplet has a wavelength of $10832.120$\,\AA, although
the authors also provide another value: $10832.150$\,\AA. 
However, when we calculate the line center position using the
Gaussian fit to the FTS spectrum we find a
wavelength of $10832.099$\,\AA. Note that the wavelengths of the
FTS atlas are not corrected for the gravity shift. This is
perfectly adequate for telluric lines, which are, indeed, not affected by it.
Other wavelength values, differing slightly from those
mentioned previously, can also be found in the literature.
These discrepancies led us to make our own estimate for the center wavelength of
the H$_2$O telluric line. The process, explained in Appendix
\ref{app:telluric}, results in a wavelength of $10832.108$\,\AA. 
This value is supported by the work of \citet{breckinridge73}, 
who inferred a wavelength of $10832.109$\,\AA\ with an accuracy approaching $\pm 1$\,m\AA.

The newly constructed wavelength array is referred to a terrestrial reference frame,
that does not account for any relative orbital motions. We followed the calibration
procedures presented in Appendix A of \citet{martinez97}, and references
therein, adapted to the Observatorio del Teide, to obtain absolute
line-of-sight velocities. Line shifts due to Earth's rotation,
orbital motion of Earth around the Sun and solar rotation have been corrected.  In
the same way, the gravitational redshift, $\Delta \lambda_\mathrm{G} =
(GM_\odot / R_\odot c^2) \lambda$ (which translates into 23 m\AA\ for
this spectral range), was also corrected in our calibration. The
effect of convective blueshift has been studied for the photospheric
\ion{Si}{i} line, however, after reviewing the literature and studying the
response function to various physical perturbations of this line, we concluded
that the correction is rather negligible owing to the formation height of the
\ion{Si}{i} line, which happens at a considerable height ($\log \tau \sim
-2$) above the surface.

\section{Determination of a new telluric line wavelength}\label{app:telluric}

As mentioned in the previous appendix, the literature quotes
several different values for the central wavelength
($\lambda_\mathrm{T}$) of the H$_2$O telluric line next to the \ion{He}{i}
triplet. We calibrated our
spectrum using this telluric line as a reference, but the resulting velocity
maps were clearly shifted to the red (when using $\lambda_\mathrm{T} =
10832.120$ \AA) or to the blue (when using $\lambda_\mathrm{T} = 10832.099$
\AA). This is, we found systematic photospheric redshifts or blueshifts (depending on
the wavelength used for the calibration) in the faculae, where velocities are expected to be around zero
\citep[see Fig.\ 12 in][where high filling factor faculae show no
velocity shift]{martinez97}. Shifts of the same order of magnitude were found when
compared to new data from a recent observing campaign in 2010 with the Tenerife
Infrared Polarimeter at the VTT. We attribute this
inconsistency to an incorrect value of the central wavelength of
the telluric line and therefore we
corrected its wavelength using the following method: we took three flat fields
from our 2010 campaign with TIP-II (August 21 and 22); two from the morning
and one from the afternoon. All of the flat fields were taken at disk center in the quiet sun,
with a random circular movement of the telescope pointing of up to 50\arcsec. These maps were then
used as input data for the standard reduction procedure including 
flat field, dark current and polarimetric calibration corrections 
\citep{collados99, collados03}. The spectral sampling for each flat field was inferred
using the same procedure described in Appendix \ref{app:a}. Using $\lambda_\mathrm{T} = 10832.120$
\AA\ as the reference telluric line, a mean redshift of the \ion{Si}{i} line of
$\Delta \lambda \sim 0.0117$ \AA\ was found. As mentioned above, 
this line is not expected to show any convective blueshift due to its 
formation height. Also no redshifts are expected at disk center. Thus, this
systematic redshift was
subtracted from the $\lambda_\mathrm{T} = 10832.120$ \AA\ telluric line, 
yielding a new wavelength of $\lambda_\mathrm{T}^\mathrm{NEW} = 10832.108$ \AA.
With the new reference, the average photospheric facular velocity for all maps
resulted in $\sim - 0.06$ km\,s$^{-1}$. Since the faculae observed here are very compact and
have large filling factors (typically, higher than 50\%), this value is expected to be zero.
We thus conclude that the systematic effects of our velocity calibration are smaller than 
60\,m\,s$^{-1}$ or around 2\,m\AA. 

\begin{acknowledgements}
This work has been partially funded by the Spanish Ministerio
de Educaci\'on y Ciencia, through Project No. AYA2009-14105-C06-03
and AYA2011-29833-C06-03. Financial support 
by the European Commission through the SOLAIRE Network (MTRN-CT-2006-035484) is gratefully acknowledged.
This paper is based on observations made with the VTT operated on the island of Tenerife by
the KIS in the Spanish Observatorio del Teide of the Instituto de Astrof\'isica
de Canarias. The National Center
for Atmospheric Research (NCAR) is sponsored by the National Science Foundation
(NSF). B. Ruiz Cobo helped with the support and implementation of the SIR code and is
gratefully acknowledged. 

\end{acknowledgements}

\bibliographystyle{aa} 
\bibliography{esquema2} 

\end{document}